\documentclass[sn-aps,referee]{sn-jnl}

\usepackage{graphicx}%
\usepackage{multirow}%
\usepackage{amsmath,amssymb,amsfonts}%
\usepackage{amsthm}%
\usepackage{mathrsfs}%
\usepackage[title]{appendix}%
\usepackage{xcolor}%
\usepackage{textcomp}%
\usepackage{manyfoot}%
\usepackage{booktabs}%
\usepackage{algorithm}%
\usepackage{algorithmicx}%
\usepackage{algpseudocode}%
\usepackage{listings}%
\theoremstyle{thmstyleone}%
\theoremstyle{thmstyletwo}%

\theoremstyle{thmstylethree}%

\begin{document}

\title[Article Title]{Closed-form expressions for the probability distribution of quantum walk on a line}








\author*[1]{\fnm{Mahesh N.} \sur{Jayakody}}\email{jadmn.jayakody@gmail.com}

\author*[1]{\fnm{Eliahu} \sur{Cohen}}\email{eliahu.cohen@biu.ac.il}


\affil[1]{\orgdiv{Faculty of Engineering and the Institute of Nanotechnology and Advanced Materials}, \orgname{Bar-Ilan University}, \orgaddress{\city{Ramat Gan}, \postcode{5290002}, \country{Israel}}}




\abstract{Theoretical and applied studies of quantum walks are abundant in quantum science and technology thanks to their relative simplicity and versatility. Here we derive closed-form expressions for the probability distribution of quantum walks on a line. The most general two-state coin operator and the most general (pure) initial state are considered in the derivation. The general coin operator includes the common choices of Hadamard, Grover, and Fourier coins. The method of Fibonacci-Horner basis for the power decomposition of a matrix is employed in the analysis. Moreover, we also consider mixed initial states and derive closed-form expression for the probability distribution of the Quantum walk on a line. To prove the accuracy of our derivations, we retrieve the simulated probability distribution of Hadamard walk on a line using our closed-form expressions. With a broader perspective in mind, we argue that our approach has the potential to serve as a helpful mathematical tool in obtaining precise analytical expressions for the time evolution of qubit-based systems in a general context.}

\keywords{Discrete-time quantum walks, Quantum walk on a line, Fibonacci-Horner power decomposition}



\maketitle

\section{Introduction}\label{sec1}

Quantum walks (QWs) are considered as the quantum analogs of classical random walks \cite{aharonov1993quantum}. Similarly to its classical counterpart, the framework of QWs has been instructive for theoretical and practical understanding of quantum algorithms \cite{kempe2003quantum, shenvi2003quantum, childs2003exponential, berry2010quantum} and quantum computing in general \cite{childs2009universal, childs2013universal}. Quantum walks have also been used in modeling transport in biological systems \cite{oliveira2006decoherence, hoyer2010limits, lloyd2011quantum} as well as physical phenomena such as Anderson localization \cite{schreiber2011decoherence, wojcik2012trapping, zhang2014one, crespi2013anderson, xue2014trapping} and topological phases \cite{kitagawa2010exploring, kitagawa2012observation}. There are two main types of quantum walks: discrete-time quantum walks (DTQWs) and continuous-time quantum walks (CTQWs). These two classes of quantum walks have distinct mathematical descriptions \cite{manouchehri2013physical}. The obvious, yet significant, difference between them is that DTQWs are defined on a discrete-time domain, while CTQWs are defined on a continuous-time domain. However, it has been demonstrated that under certain conditions, DTQWs can be transformed into CTQWs \cite{Strauch_2006}. In this paper, we specifically analyze DTQWs on the integer line. These walks are often referred to as one dimensional discrete-time QWs or simply as QWs on a line, which possess the simplest structure among QWs, are nevertheless useful for understanding higher-dimensional walks on graphs \cite{qw_graphs_2002}. Given the wide range of applications associated with QWs, it is indeed important to have analytical knowledge of the probability distribution of the quantum walker in order to fully understand and utilize the quantum advantage offered by framework of QW. The probability distribution provides information about the likelihood of the quantum walker being at different positions during the walk. Analytical expressions for the probability distribution allow for a deeper understanding of the behavior of the walker and facilitate the design and analysis of quantum algorithms, simulations, and other applications. It is worth noting that obtaining explicit analytical expressions for the probability distribution can be challenging, especially for more complex QWs models. In such cases, numerical simulations or approximation techniques are often used to study the behavior of QWs \cite{venegas2012quantum}. 

Several analytical studies can be found in the literature which adopt different approaches for deriving explicit expressions for the probability distributions associated to QWs on a line \cite{lavivcka2011quantum}. Discrete time Fourier transform as well as methods from complex analysis have been used by Nayak and Vishwanath in \cite{nayak2000quantum} to launch an analytical study on one-dimensional Hadamard walk. Moreover, in that study, Nayak and Vishwanath determine the asymptotic probability distribution of Hadamard walk on a line for a specific initial state. In \cite{kovsik2003two}, a closed-form expression is given for the probability of finding the quantum walker at the origin after $n$ steps of the Hadamard walk on a line when starting from a localized initial state. Another approach based on the path integral formulation of quantum mechanics is adopted in \cite{ambainis2001one} for deriving an asymptotic probability distribution of the Hadamard walk on a line. In that approach, the coin coefficients of the quantum walker are expressed in terms of Jacobi polynomials and then the probability distribution of the walker is derived by determining the asymptotic behaviour of these Jacobi polynomials. A comprehensive review of the path integral and Schr\"odinger approaches, which are used to study the Hadamard walk analytically, is given in \cite{carteret2003three} along with a new way of analyzing the discrete-time QW on the infinite line in terms of Airy functions. Fuss et al. \cite{fuss2007analytic} have introduced an analytical framework to describe the probability densities and moments of one-dimensional QWs on a line. Additionally, an alternative formulation of DTQWs, rooted in scattering theory, was proposed by Feldman and Hillery \cite{feldman2007modifying}. The limiting distribution of the QW on a line for a special delocalized initial state is derived analytically in \cite{machida2013quantum}. An asymptotic probability distribution of the QW on a line with the most general coin operator and the most general localized initial state is derived in \cite{konno2005new}. 

Most of the aforementioned studies provide exact probability distributions of QWs on a line in the long time limit (i.e. asymptotic probability distributions). Nonetheless, closed-form expressions of the probability distribution derived for a finite time step could be useful too when developing QW-based algorithmic applications and simulations. Hence, in this paper, we derive closed-form expressions, associated with the probability distribution of the QW on a line, for an arbitrary time step $t$. The most general two-state coin operator and the most general (pure) initial state are considered in the derivation which is based on the Fibonacci-Horner approach for the power decomposition of a matrix. Moreover, we also consider mixed initial states and derive closed-form expressions for the probability distribution of the Hadamard walk on a line. The paper is organized as follows. In Sec. \ref{QW on a line}, the mathematical framework of QWs on a line is given. The time evolution of the QW is formulated in Sec. \ref{time evolution of QWs}. We dedicate Sec. \ref{Prob Dist of QWs} and Sec. \ref{Mixed initial states} to explain the derivation of exact expressions related to the most general case and to a specific case of QWs on a line, respectively. Finally, to prove the accuracy of our derivations, we retrieve the simulated probability distribution of Hadamard walk on a line using our closed-form expressions.

\section{Quantum Walks on a line}\label{QW on a line}
Let us consider the standard model of QW on a line which comprises a two-state coin and a walker. The evolution of the walker on the one-dimensional integer line is conditioned upon the state of the coin outcome. The Hilbert space associated with the joint system is denoted by $\mathcal{H}_w \otimes \mathcal{H}_c$, where $\mathcal{H}_w$ is the space of the walker and $\mathcal{H}_c$ is the space of the coin. A single-step progression of the system consists of a transformation applied to the coin system followed by a conditional shift of the walker (the walker moves either to the left or right upon the condition of the coin outcome). We write the unitary operator corresponding to this single-step evolution as $U=SC$, where $S$ and $C$ are the shift and coin-flip operators, respectively. Here, we consider the conventional shift operator, defined by
\begin{equation}\label{eq:1d-s}
   S=  \sum_{x=-\infty}^{\infty}{|x+1\rangle\langle x|} \otimes |0\rangle \langle 0|+ |x-1\rangle\langle x| \otimes |1\rangle \langle 1|
\end{equation}
where, as usual, $|0\rangle \langle 0|=\left (\begin{array}{cc}
 1 & 0 \\
 0 & 0 \\
\end{array}
\right)$ and $|1\rangle \langle 1|=\left (\begin{array}{cc}
 0 & 0 \\
 0 & 1 \\
\end{array}
\right)$. In addition, we consider the most general two-state coin operator \cite{manouchehri2013physical} of the QW on a line which can be expressed as
\begin{equation}\label{coin_matrix}
C=
\left( {\begin{array}{cc}
\cos \theta & e^{i\phi_1}\sin \theta \\
e^{i\phi_2}\sin \theta & -e^{i(\phi_1+\phi_2)}\cos \theta \\
\end{array}}\right)
 \end{equation}
where $\theta\in [0,2\pi)$ and $\phi_1,\phi_2\in [0,\pi)$. Note that, when $\theta=\frac{\pi}{4}$ and $\phi_1=\phi_2=0$, the coin $C$ becomes the common Hadamard coin operator. Moreover, for the condition of $\theta=\pi$ and $\phi_1=\phi_2=0$, we obtain the Grover coin. The combination of $\theta=\frac{\pi}{4}$ and $\phi_1=\phi_2=\pi$, yields the Fourier coin \cite{Gratsea}. The initial state of the coin and position of the quantum walker play a significant role in QWs compared to that of a classical random walk. It is a well-known fact that, for a given QW, different initial states yield different probability distributions. This behaviour is unique to QWs and has no exact analogy within classical random walks. The most general (pure) initial state of the quantum walker can be written as
 \begin{equation}\label{initial_state}
     |\psi_0\rangle =\sum_{x'=-n}^{n}{|x'\rangle \otimes \bigg(\alpha_{x'}| 0 \rangle+\beta_{x'} |1 \rangle \bigg)}
 \end{equation}
 where $n \in \mathbb{N}$ and for each $x'$, the coefficients $\alpha_{x'}, \beta_{x'} $ are complex numbers that hold the relationship of $\sum_{x'}{|\alpha_{x'}|}^2+{|\beta_{x'}|}^2=1$. The physical meaning of the expression given in \eqref{initial_state} is that, at the beginning, the quantum walker potentially placed at all the locations from $x'=-n$ to $x'=n$. In other words, this is, in general, a superposition of all position states. Moreover, at each location $x' \in [-n ,n]$, the walker holds a superposition of coin states given by $\alpha_{x'}| 0 \rangle+\beta_{x'} |1 \rangle$. When the quantum walker has non-zero coin coefficients in more than two $x'$ locations (i.e. $\alpha_{x'} \neq 0$ or/and $\beta_{x'} \neq 0$ ), the initial state given in \eqref{initial_state} is termed ``delocalized''. Moreover, when the walker is placed only in a single location, the initial state is termed ``localized''. 

\section{Time evolution of QWs} \label{time evolution of QWs}
 A single time-step evolution of the coin-walker system is determined by applying the unitary operator $U$ on the present state of the quantum walker. Hence, the time evolution of the quantum walker can be written as
\begin{equation}\label{time evolution}
     |\psi_t\rangle =U^{t}|\psi_0\rangle
\end{equation}
where $|\psi_0\rangle$ is the initial state and $|\psi_t\rangle$ is the state of the quantum walker after $t$ time steps. The explicit form of $|\psi_t\rangle$ is given by
\begin{equation}\label{state_at_t}
     |\psi_t\rangle =\sum_x{|x\rangle \otimes \bigg(\alpha_x(t) | 0 \rangle+\beta_x (t) |1 \rangle \bigg)}
\end{equation}
where $t \in \mathbb{Z^+}$ and the coefficients $\alpha_x(t) , \beta_x(t) \in \mathbb{C}$ holds the relationship of $\sum_x{{|\alpha_x(t)|}^2+{|\beta_x(t)|}^2=1}$. For the sake of convenience, let us conduct our analysis in the momentum space. The transformation rule from the $x$-basis to the $k$-basis is given by 
\begin{equation}
     |k\rangle =\sum_x{e^{ikx}}|x\rangle,
\end{equation}
where $k \in \left[-\pi,\pi\right]$. The inverse transformation is given by
\begin{equation}
    |x\rangle =\frac{1}{2\pi}\int_{-\pi}^{\pi}dk \ e^{-ikx}|k\rangle,
\end{equation}
Thus, in momentum space, $U$ is represented by the unitary operator $U_k$ given by
\begin{equation}
    U_k=\left( {\begin{array}{cc}
e^{-ik}\cos \theta & e^{-ik}e^{i\phi_1}\sin \theta \\
e^{ik}e^{i\phi_2}\sin \theta & -e^{ik}e^{i(\phi_1+\phi_2)}\cos \theta \\
\end{array}}\right)
\end{equation}
 This representation allows us to view a single-step evolution of an arbitrary state of the coin-walker system as a transformation performed only on the coin state of that arbitrary state. Following the analysis given in \cite{abal2006quantum,nayak2000quantum} we can transform the state vector in \eqref{state_at_t} to the momentum space. Then, the transformed coefficients of the coin state in \eqref{state_at_t} are given by
 \begin{equation}\label{coefficients in k basis}
 \begin{split}
     \tilde{\alpha}_k(t)& = \sum_x e^{ikx}\alpha_x(t)\\
      \tilde{\beta}_k(t)& = \sum_x e^{ikx}\beta_x(t)\\
 \end{split}
\end{equation}
The inverse transformation is given by
 \begin{equation}\label{coefficients in x basis}
 \begin{split}
    \alpha_x(t) &= \frac{1}{2\pi}\int_{-\pi}^{\pi}e^{-ikx}\tilde{\alpha}_k(t)dk \\
    \beta_x(t) &= \frac{1}{2\pi}\int_{-\pi}^{\pi}e^{-ikx}\tilde{\beta}_k(t)dk
 \end{split}
\end{equation}
A single step evolution of the coin-walker system in $k$-basis can be written as 
\begin{equation}\label{k basis single step evolution}
 \begin{pmatrix}
     \tilde{\alpha}_k(t+1) \\
     \tilde{\beta}_k(t+1)
 \end{pmatrix}=U_k\begin{pmatrix}
     \tilde{\alpha}_k(t) \\
     \tilde{\beta}_k(t)
 \end{pmatrix}
\end{equation}
Hence, the time evolution of the quantum walker can be formulated as
 \begin{equation}\label{time evolution k basis}
     \begin{pmatrix}
     \tilde{\alpha}_k(t) \\
     \tilde{\beta}_k(t)
 \end{pmatrix}=U_k^{t}\begin{pmatrix}
     \tilde{\alpha}_k(0) \\
     \tilde{\beta}_k(0)
 \end{pmatrix}
 \end{equation}
 where $(\tilde{\alpha}_k(0),\tilde{\beta}_k(0))^T$ is the initial state of the quantum walker in momentum space. One can determine the coefficients of the initial state in momentum space by applying the transformation formulas in \eqref{coefficients in k basis} to \eqref{initial_state}. Then, the transformed coefficients of the initial state can be written as
\begin{equation}\label{initial state in momentum space}
\begin{split}
\tilde{\alpha}_k(0)& = \sum_{x'} e^{ikx'}\alpha_{x'}\\
\tilde{\beta}_k(0)& = \sum_{x'} e^{ikx'}\beta_{x'}\\
\end{split}
\end{equation}
where the coefficients $\alpha_{x'}, \beta_{x'} $ are complex numbers that hold the relationship of $\sum_{x'}{|\alpha_{x'}|}^2+{|\beta_{x'}|}^2=1$. 
\section{Probability Distribution} \label{Prob Dist of QWs}
\begin{figure}[!ht]
\centerline{\includegraphics[width=3.0in, height=2.0in]{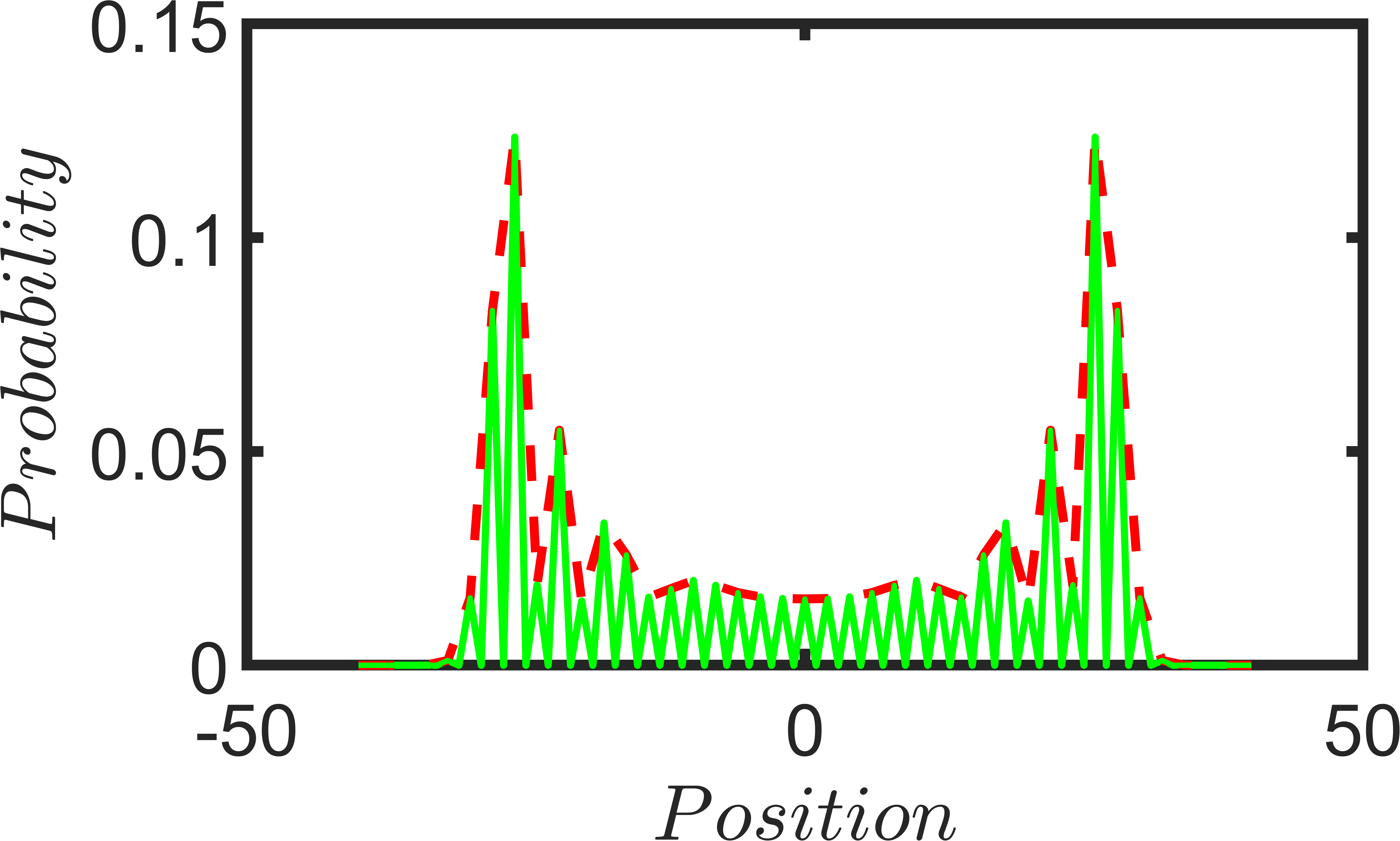}}
\caption {Probability distribution of the Hadamard walk initiated from the localized state of $|0 \rangle \otimes \frac{1}{\sqrt{2}}(|0\rangle + i|1\rangle)$ and evolved for $40$ time steps. (a) The dashed line (red) shows the theoretical distribution. (b) The solid line (green) shows the simulated probability distribution. Since the time step is an even number, the walker resides only on even sites after 40 time steps.}
\label{fig1}
\end{figure}

The probability distribution of the QW on a line describes the likelihood of finding the walker at different locations along the integer line after a certain number of steps. From \eqref{state_at_t}, the probability $P(x,t)$ of finding the quantum walker at position $x$ at time $t$ can be determined by taking the absolute square of the coin coefficients of the walker at $x$. Hence, the explicit form of $P(x,t)$ can be written as  
\begin{equation}\label{probability Eq1}
P(x,t)=|\alpha_x(t)|^2+|\beta_x(t)|^2
\end{equation}
where $\alpha_x(t) , \beta_x(t) \in \mathbb{C}$. Thus, to derive an analytic expression for the probability distribution of the QW on a line, we need to seek explicit expressions for the coin coefficients of $\alpha_x(t)$ and $\beta_x(t)$. According to the relationship given in \eqref{time evolution k basis}, if one can determine an analytical form for the $t^{th}$ power of the $U_k$ operator, then explicit expressions for the coin coefficients at time $t$ can be determined by applying $U_k^{t}$ on the initial state. The standard practice of determining the $t^{th}$ power of $U_k$ involves expressing the operator $U_k$ in its eigen decomposition and then calculating the $t^{th}$ power based on the eigenvalues and eigenvectors. However, the result obtained from the eigen decomposition of $U_k$ gives rise to complex functions which are difficult to integrate over $k$ when taking the inverse transformation. Therefore, to avoid this, we utilize the Fibonacci-Horner power decomposition \cite{laarichi2022explicit} to find the $t^{th}$ power of the $U_k$ operator.  The Fibonacci-Horner power decomposition uses the coefficients of the characteristic polynomial of a matrix to calculate its powers. The characteristic equation of $U_k$ is given by
\begin{equation}\label{characteristic equation of U_k }
R(\lambda)=\lambda^2-c_0\lambda-c_1
\end{equation}
where $c_0=(e^{-ik}-e^{ik})e^{i(\phi_1+\phi_2)}\cos{\theta}$ and $c_1=e^{i(\phi_1+\phi_2)}$. Then, the $t^{th}$ power of $U_k$ can be expressed as follows
\begin{equation}\label{power of Uk}
U^t_k=f_t\mathbb{I}+f_{t-1}(U_{k}-c_0\mathbb{I})
\end{equation}
where $\mathbb{I}$ is the $2 \times 2$ identity matrix and the function $f_t$ is given by
\begin{equation}\label{function f}
 f_t=\displaystyle{\sum_{h_0+2h_1=t}{\frac{(h_0+h_1)!}{h_0!h_1!} c_{0}^{h_0}c_{1}^{h_1}}}
\end{equation}
with $c_0=(e^{-ik}-e^{ik})e^{i(\phi_1+\phi_2)}\cos{\theta}$ and $c_1=e^{i(\phi_1+\phi_2)}$. By substituting \eqref{power of Uk} in \eqref{time evolution k basis} we have the following pair of relations
\begin{equation}\label{expanded coeff in k basis}
\begin{split}
\tilde{\alpha}_k(t)&=f_t\tilde{\alpha}_k(0) +f_{t-1}e^{i(k+\phi_1 +\phi_2)}\cos \theta \tilde{\alpha}_k(0) +f_{t-1}e^{-i(k-\phi_1)}\sin\theta\tilde{\beta}_k(0) \\
\tilde{\beta}_k(t)&=f_t\tilde{\beta}_k(0) -f_{t-1}e^{-ik}\cos \theta \tilde{\beta}_k(0) +f_{t-1}e^{i(k+\phi_2)}\sin\theta\tilde{\alpha}_k(0)\\
\end{split}
\end{equation}
Note that one can determine the coin coefficients in the $x$ basis by applying the inverse transformation formulas to \eqref{expanded coeff in k basis}. However, the exponential terms, namely, the function $f_t$ as well as the coefficients $\tilde{\alpha}_k(0)$ and $\tilde{\beta}_k(0)$ of the initial state, contain the variable $k$. Hence, when taking the integral over $k$, one needs to take all the aforementioned terms into consideration. The detailed calculation is given in Appendix \ref{explict form of coin coeff}. The closed-form expressions for the coin coefficients of the QW with the most general coin operator and the most general (pure) initial state, which are given in \eqref{coin_matrix} and \eqref{initial_state} respectively, can be written as follows

\begin{equation}\label{coin coeff in x-basis for localized state}
    \begin{split}
        \alpha_x(t)&=\sum_{h=0}^t \sum_{x'} (-1)^{\frac{h-x'+x}{2}}\frac{(\frac{t+h}{2})!}{(\frac{t-h}{2})!(\frac{h+x'-x}{2})!(\frac{h-x'+x}{2})!}(\cos\theta)^{h} \chi^{\frac{t-x'+x}{2}}\alpha_{x'} \\
        & +\sum_{h=0}^{t-1} \sum_{x'} (-1)^{\frac{h-1-x'+x}{2}}\frac{(\frac{t-1+h}{2})!}{(\frac{t-1-h}{2})!(\frac{h+1+x'-x}{2})!(\frac{h-1-x'+x}{2})!}(\cos\theta)^{h+1}\chi^{\frac{t-x'+x}{2}}\alpha_{x'} \\
        & +\sum_{h=0}^{t-1}\sum_{x'} (-1)^{\frac{h+1-x'+x}{2}}\frac{(\frac{t-1+h}{2})!}{(\frac{t-1-h}{2})!(\frac{h-1+x'-x}{2})!(\frac{h+1-x'+x}{2})!}(\cos\theta)^{h}\sin\theta \ \chi^{\frac{t-x'+x}{2}} e^{i\phi_1} \beta_{x'} \\
        \beta_x(t)&=\sum_{h=0}^t \sum_{x'}(-1)^{\frac{h-x'+x}{2}}\frac{(\frac{t+h}{2})!}{(\frac{t-h}{2})!(\frac{h+x'-x}{2})!(\frac{h-x'+x}{2})!}(\cos\theta)^{h}\chi^{\frac{t-x'+x}{2}}\beta_{x'} \\
        & + \sum_{h=0}^{t-1}\sum_{x'} (-1)^{\frac{h-1-x'+x}{2}}\frac{(\frac{t-1+h}{2})!}{(\frac{t-1-h}{2})!(\frac{h+1+x'-x}{2})!(\frac{h-1-x'+x}{2})!}(\cos\theta)^{h} \sin \theta  \chi^{\frac{t-x'+x}{2}} e^{-i\phi_1} \alpha_{x'} \\
        & - \sum_{h=0}^{t-1}\sum_{x'}(-1)^{\frac{h+1-x'+x}{2}}\frac{(\frac{t-1+h}{2})!}{(\frac{t-1-h}{2})!(\frac{h-1+x'-x}{2})!(\frac{h+1-x'+x}{2})!}(\cos\theta)^{h+1}\chi^{\frac{t-x'+x}{2}} \beta_{x'}
    \end{split}
\end{equation}
where $\chi=e^{i(\phi_1 +\phi_2)}$. Figure \ref{fig1} shows the theoretical and simulated probability distributions of the Hadamard walk initiated from a pure localized state. According to the simulated probability distribution in Figure \ref{fig1}, the walker resides only on even sites after 40 time steps. A similar behavior can be observed in the theoretical distribution as well. However, for the sake of clarity, we have excluded the sites with zero probability when plotting the theoretical distribution. It is important to emphasize that the factorial terms in \eqref{coin coeff in x-basis for localized state} impose constraints on the sites which can be occupied by the walker. Hence, by analyzing the factorial terms, one can determine the allowed/forbidden sites for the walker at a given time step. Accordingly, the theoretical distribution in Figure \ref{fig1} dictates that the Hadamard walker is allowed only to occupy even sites after 40 time steps.

\section{Mixed initial states}\label{Mixed initial states}
The most general initial state given in \eqref{initial_state} is a pure state. Nonetheless, QWs can also be initiated from a mixed initial state as well. Here, what we mean by ``mixed'' is a mixed state in coin space. However, one can define mixed states in position space or in both coin and position spaces as well. During the quantum state preparation procedure that yields the initial state of the QW, it is possible to have a classical mixture of initial states due to the imperfection of the procedure. For instance, consider a scenario in which the initial state of $|0\rangle \otimes |0 \rangle$ is prepared with $\mu$ probability and the initial state of $|0\rangle \otimes |1 \rangle$ is prepared with $1-\mu$ probability. A few studies can be found in literature where mixed initial states are considered in the context of QWs \cite{kollar2012asymptotic, kollar2014entropy,tregenna2003controlling,jeong2013experimental}. In this section, we consider mixed initial states in coin space while keeping the initial position localized at the origin. We henceforth call such states ``mixed localized initial states''. When dealing with QWs that start with an ensemble as the initial state rather than a pure state, the density operator formalism is required. Hence, in this section, we adopt the density operator formalism for our analysis. The density operator that corresponds to the most general mixed localized initial state of a QW on a line can be written as
\begin{equation}\label{mixed initial state}
    \rho_0=|0\rangle\langle 0| \otimes \left (\begin{array}{cc}
 \rho_{11} & \rho_{12} \\
 \rho_{21} & \rho_{22} \\
\end{array}
\right)
\end{equation}
where $\rho_{ij} \in \mathbb{C}$. In the momentum representation, ${\rho }_0$ takes the following from 
\begin{equation}\label{Eq:C5_34}
 {\rho }_0=\int^{\pi }_{-\pi }{\frac{dk}{2\pi }} \int^{\pi }_{-\pi }{\frac{dk'}{2\pi }}|k\rangle \langle k'| \otimes \mathbb{O}
\end{equation}
where $\mathbb{O}=\left (\begin{array}{cc}
 \rho_{11} & \rho_{12} \\
 \rho_{21} & \rho_{22} \\
\end{array}
\right)$. Following the analysis given in \cite{jayakody2021analysis}, the state after time $t$ can be written using a super-operator $\mathcal{L}_{k,k'} $ as follows
\begin{equation}\label{Eq:C5_35}
 {\rho }_t=\int^{\pi }_{-\pi }{\frac{dk}{2\pi }} \int^{\pi }_{-\pi }{\frac{dk'}{2\pi }}|k\rangle \langle k'| \otimes {\mathcal{L}}^t_{k,k'}\mathbb{O}
\end{equation} 
where ${\mathcal{L}}_{k,k'} \mathbb{O} =U_k\mathbb{O}{U_{k'}}^{\dagger }$. The position density matrix can be obtained by tracing out the coin space from \eqref{Eq:C5_35}. Then, the expression for the position density matrix can be written as
\begin{equation}\label{position_density_matrix}
 {\rho }_w(t)=\int^{\pi }_{-\pi }{\frac{dk}{2\pi }} \int^{\pi }_{-\pi }{\frac{dk'}{2\pi }}|k\rangle \langle k'| \times Tr\bigg( {\mathcal{L}}^t_{k,k'} \mathbb{O}\bigg)
\end{equation} 
The probability of finding the particle at position $y$ at time $t$ can be determined by
\begin{equation}\label{position_probability}
    P(y,t)=\langle y |{\rho }_w(t)|y\rangle=\int^{\pi }_{-\pi }{\frac{dk}{2\pi }} \int^{\pi }_{-\pi }{\frac{dk'}{2\pi }}e^{iy(k-k')}Tr\bigg( {\mathcal{L}}^t_{k,k'} \mathbb{O}\bigg)
\end{equation} 
Now, our next task is to derive an analytical expression for the trace of the operator ${\mathcal{L}}^t_{k,k'} \mathbb{O}$. To make the computation much easier, we fix the coin operator to the $2 \times 2$ Hadamard coin. The matrix form of the $2 \times 2$ Hadamard coin operator is given by
\begin{equation}\label{Hadmard coin}
H=
\frac{1}{\sqrt{2}}\left( {\begin{array}{cc}
1 & 1 \\
1 & -1 \\
\end{array}}\right),
 \end{equation}
Note that a convenient representation of $\mathbb{O}$ can be written as  
\begin{equation}\label{Eq:C5_22} 
 \mathbb{O}=r_0{\sigma }_0+r_1{\sigma }_1+r_2{\sigma }_2+r_3{\sigma }_3
\end{equation}
where $r_i \in \mathbb{R}$, ${\sigma }_0=\mathbb{I}$ and ${\sigma }_{1,2,3}={\sigma }_{x,y,z}$ are the usual Pauli matrices. Then, the block matrix for ${\mathcal{L}}_{k,k'}\mathbb{O}$ can be written in the Pauli basis as 
\begin{equation}\label{sup_Operator}
{\mathcal{L}}_{k,k'}\mathbb{O}=\left({\begin{array}{cccc}
 L_{11} & L_{12} & 0 & 0 \\
 0 & 0 & L_{23} & L_{24} \\
 0 & 0 & L_{33} & L_{34} \\
 L_{41} & L_{42} & 0 & 0 \\
\end{array}}\right)\left({\begin{array}{c}
 r_0 \\
 r_1 \\
 r_2 \\
 r_3 \\
\end{array}}\right ),
\end{equation}
where $L_{11} = \cos{(k - k')} \\
L_{12}= -i \sin{(k - k')} \\
L_{23} = \sin{(k + k')} \\
L_{24} = \cos{(k + k')}  \\
L_{33} = -\cos{(k + k')} \\
L_{34} = \sin{(k + k')}  \\
L_{41} = -i\sin{(k - k')} \\
L_{42} = \cos{(k - k')}$ \\

The $t^{th}$ power of the block matrix corresponding to the super-operator ${\mathcal{L}}_{k,k'}$ can be determined by using the Fibonacci-Horner power decomposition. The detailed calculation of determining the $t^{th}$ power is given in Appendix \ref{Power of L_kk}. 

\begin{figure}[!ht]
\centerline{\includegraphics[width=2.5in, height=1.5in]{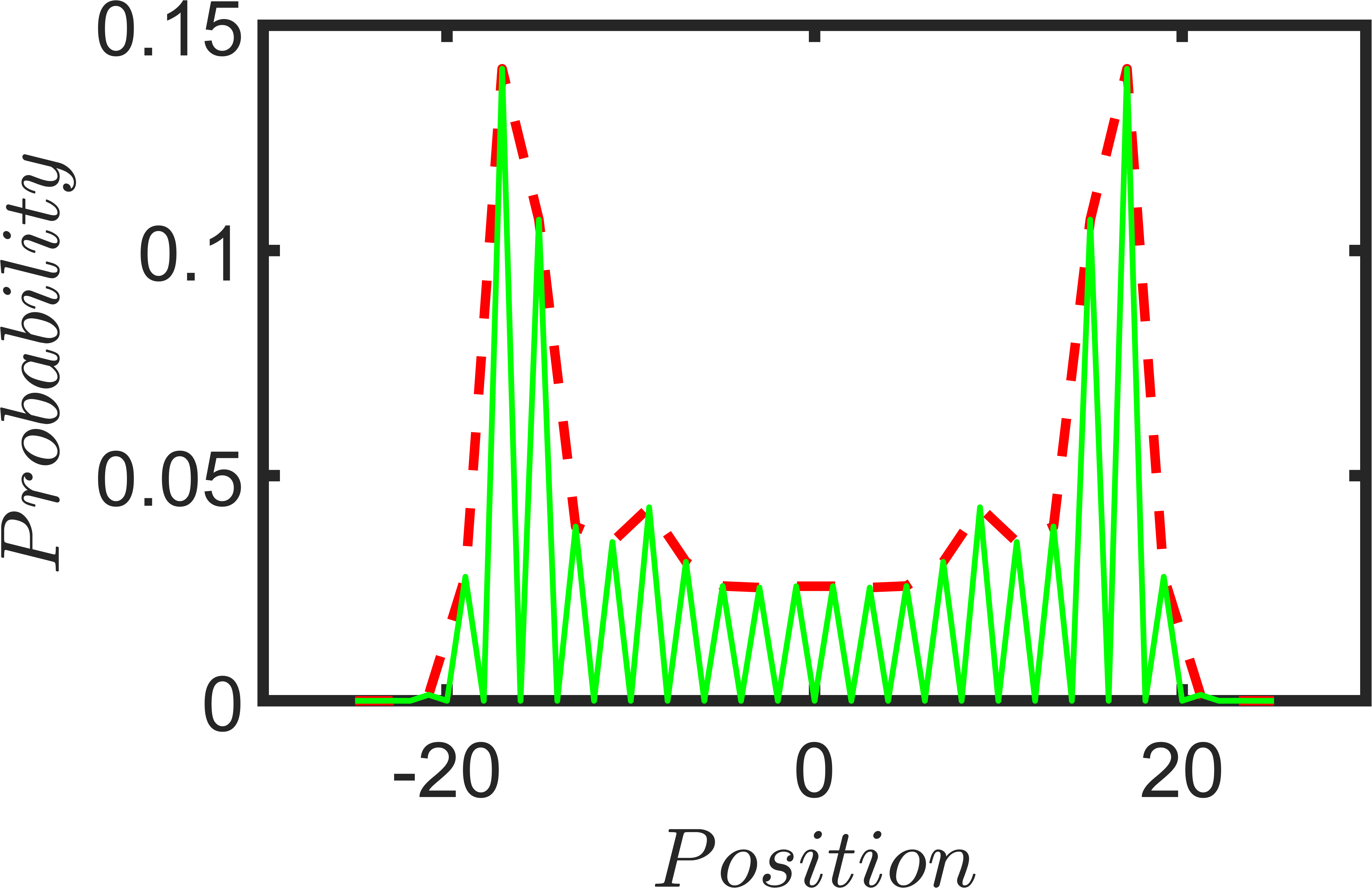}}
\caption {Probability distribution of the Hadamard walk that initiates from the mixed localized initial state $|0\rangle\langle 0| \otimes \frac{1}{2}\left (\begin{array}{cc}
 1 & 0 \\
 0 & 1 \\
\end{array}
\right)$. This corresponds to the scenario in which the initial states of $|0\rangle \otimes |0 \rangle$ and $|0\rangle \otimes |1 \rangle$ are prepared with $\frac{1}{2}$ probability. The QW propagates for $25$ time steps. (a) The dashed line (red) shows the theoretical distribution. (b) The solid line (green) shows the simulated probability distribution. Since the time step is an odd number, the walker resides only on odd sites after 25 time steps.}
\label{fig2}
\end{figure}

Now let us calculate the probability distribution related to the mixed localized initial state. For that, we must determine the trace of  ${\mathcal{L}}_{k,k'}^{t}\mathbb{O}$ and then evaluate the integral in \eqref{position_probability}. The detailed calculation is given in Appendix \ref{Prob_dis_of_mixed_localized_state}. The probability distribution $P_{l}(y,t)$ related to the mixed localized initial state is given by
\begin{equation}
\begin{split}\label{Mixed_Prob_dis}
    & P_{l}(y,t)=2r_0 \sum_{\overline{h}=t} \sum_{s_0=0}^{h_0}\sum_{s_2=0}^{h_2} A_0 \begin{pmatrix} A_1 \\ \frac{A_1-y}{2}\\ \end{pmatrix} \begin{pmatrix} A_2 \\ \frac{A_2}{2}\\ \end{pmatrix} \\
    & +  \sum_{\overline{h}=t-1} \sum_{s_0=0}^{h_0}\sum_{s_2=0}^{h_2} A_0 \bigg[2r_0\begin{pmatrix} A_1 \\ \frac{A_1-y}{2}\\ \end{pmatrix} \begin{pmatrix} A_2 \\ \frac{A_2-1}{2}\\ \end{pmatrix} + \frac{r_2y}{A_1+1}\begin{pmatrix} A_1+1 \\ \frac{A_1-y+1}{2}\\ \end{pmatrix} \begin{pmatrix} A_2 \\ \frac{A_2}{2}\\ \end{pmatrix}\bigg] \\
    & -r_0\sum_{\overline{h}=t-2} \sum_{s_0=0}^{h_0}\sum_{s_2=0}^{h_2} A_0 \begin{pmatrix} A_1+1 \\ \frac{A_1-y+1}{2}\\ \end{pmatrix} \begin{pmatrix} A_2 \\ \frac{A_2-1}{2}\\ \end{pmatrix} \\
    &- \sum_{\overline{h}=t-3} \sum_{s_0=0}^{h_0}\sum_{s_2=0}^{h_2} A_0 \bigg[r_0\begin{pmatrix} A_1+1 \\ \frac{A_1-y+1}{2}\\ \end{pmatrix} \begin{pmatrix} A_2 \\ \frac{A_2}{2}\\ \end{pmatrix} - \frac{r_3y}{A_1+1}\begin{pmatrix} A_1+1 \\ \frac{A_1-y+1}{2}\\ \end{pmatrix} \begin{pmatrix} A_2 \\ \frac{A_2}{2}\\ \end{pmatrix}\bigg] \\
\end{split}
\end{equation}
where $\begin{pmatrix} n \\ m\\
\end{pmatrix}$ is the binomial coefficient and $\overline{h} = h_0+2h_1+3h_2+4h_3$. The terms $A_1$ and $A_2$ which generate the binomial coefficients are given by $A_1=h_1+s_0+s_2$ and $A_2=h_0+h_1+h_2-s_0-s_2$. Moreover, the coefficient $A_0$ is given by $A_0=\frac{2^{h_1}}{2^{(A_1+A_2)} }\bigg[(-1)^{h_0+h_2+h_3-s_0-s_2}\bigg]\bigg[\frac{(h_0+h_1+h_2+h_3)!}{h_0!h_1!h_2!h_3!}\bigg]\begin{pmatrix} h_0 \\ s_0\\ \end{pmatrix} \begin{pmatrix} h_2 \\ s_2\\ \end{pmatrix}$. Figure \ref{fig2} shows the theoretical and simulated probability distributions of the Hadamard walk that initiates from an unbiased mixed localized initial state. According to the simulated probability distribution in Figure \ref{fig2}, the walker resides only on odd sites after 25 time steps. A similar behavior can be observed in the theoretical distribution as well. However, for the sake of clarity, we have excluded the sites with zero probability when plotting the theoretical distribution. It is important to emphasize that the factorial terms in \eqref{Mixed_Prob_dis} impose constraints on the sites which can be occupied by the walker. Hence, by analyzing the factorial terms, one can determine the allowed/forbidden sites for the walker at a given time step. Accordingly, the theoretical distribution in Figure \ref{fig2} dictates that the Hadamard walker is allowed only to occupy odd sites after 25 time steps.

\section{Conclusion}
In this paper we derive closed-form expressions associated with the probability distribution of QWs on a line for an arbitrary time step $t$. We consider the most general case of the QW on a line which comprises the most general $2 \times 2$ coin operator and the most general (pure) initial state. In addition, as a special case, we consider the Hadamard walk on a line that commences from a mixed localized initial state and derive exact expressions for the corresponding probability distribution. Unlike asymptotic distributions of the QW which are valid in the long time limit, our expressions are valid for any time step $t$, and could be useful for gaining a deeper understanding of the behaviour of the quantum walker when designing QW-based algorithms and simulations. One possible future extension of this study could be generalization of our closed-form expression by including cyclic QWs as well as higher-dimensional QWs. Transforming the unitary operator of the QW to the momentum space and determining the $t^{th}$ power of it using the Fibonacci-Horner power decomposition may pave a way for such a generalization. Moreover, the mathematical tools used in this paper may help to derive analytical expressions for the probability distribution of QWs including some decoherence parameters in certain noisy scenarios. Such an attempt would be beneficial in enhancing the comprehension of decoherence effects and would assist to mitigate decoherence in QW based algorithms and simulations.

\backmatter

\bmhead{Acknowledgments}
This research was funded by the Israel Innovation authority under grant No. 73795, by the Israeli Ministry of Science and Technology under grant No. 207059, by the Pazy foundation, by the Quantum Science and Technology Program of the Israeli Council of Higher Education and by the President Scholarship Program at Bar-Ilan University.

\section*{Author contribution}
All authors contributed equally to the paper.

\section*{Data Availability Statement}
This manuscript has no associated data or the data will not be deposited.






\begin{appendices}

\section{Derivation of the explicit form of coin coefficients $\alpha_x(t)$ and $\beta_x(t)$} \label{explict form of coin coeff}
\noindent Let us substitute the coefficients of $c_0$ and $c_1$ and rewrite $f_t$ in \eqref{function f} as follows
\begin{equation}\label{function f 2}
 f_t=\displaystyle{\sum_{h=0}^t\sum_{g=0}^{h}(-1)^{h-g}{\frac{(\frac{t+h}{2})!}{h!(\frac{t-h}{2})!}}(\cos \theta)^{h}\begin{pmatrix}
     h \\
     g \\
 \end{pmatrix} \chi^{\frac{t+h-2g}{2}} e^{ik(h-2g)}}
\end{equation}
where $\begin{pmatrix}
     h \\
     g \\
 \end{pmatrix}$ is the binomial coefficient and $\chi=e^{i(\phi_1 +\phi_2)}$. Note that here $f_t$ is written as an explicit function of variable $k$. By substituting \eqref{function f 2} and \eqref{initial state in momentum space} in \eqref{expanded coeff in k basis} and taking the inverse transformation, we can write the coin coefficients corresponding to position $x$ at time $t$ as follows
\begin{equation}\label{Eq 2}
\begin{split}
\alpha_x(t)&=\sum_{h=0}^t \sum_{g=0}^{h}\sum_{x'}(-1)^{h-g}\frac{(\frac{t+h}{2})!}{h!(\frac{t-h}{2})!}(\cos\theta)^{h}\chi^{\frac{t+h-2g}{2}}\alpha_{x'} \begin{pmatrix}
     h \\
     g \\
 \end{pmatrix}\int_{-\pi}^{\pi}\frac{dk}{2\pi}e^{ik(h-2g+x'-x)} \\
    & +\sum_{h=0}^{t-1}\sum_{g=0}^{h}\sum_{x'}(-1)^{h-g}\frac{(\frac{t-1+h}{2})!}{h!(\frac{t-1-h}{2})!}(\cos\theta)^{h+1}\chi^{\frac{t+1+h-2g}{2}}\alpha_{x'} \begin{pmatrix}
     h \\
     g \\
 \end{pmatrix}\int_{-\pi}^{\pi}\frac{dk}{2\pi}e^{ik(h-2g+1+x'-x)}\\
    & +\sum_{h=0}^{t-1}\sum_{g=0}^{h}\sum_{x'}(-1)^{h-g}\frac{(\frac{t-1+h}{2})!}{h!(\frac{t-1-h}{2})!}(\cos\theta)^{h}\sin\theta \ \chi^{\frac{t-1+h-2g}{2}}e^{i\phi_1}\beta_{x'} 
    \begin{pmatrix}
     h \\
     g \\
 \end{pmatrix}\int_{-\pi}^{\pi}\frac{dk}{2\pi}e^{ik(h-2g-1+x'-x)}\\
\beta_x(t)&=\sum_{h=0}^t \sum_{g=0}^{h} \sum_{x'}(-1)^{h-g}\frac{(\frac{t+h}{2})!}{h!(\frac{t-h}{2})!}(\cos\theta)^{h} \begin{pmatrix}
     h \\
     g \\
 \end{pmatrix}\chi^{\frac{t+h-2g}{2}} \beta _{x'} \int_{-\pi}^{\pi}\frac{dk}{2\pi}e^{ik(h-2g+x'-x)}\\
        &+ \sum_{h=0}^{t-1}\sum_{g=0}^{h} \sum_{x'}(-1)^{h-g}\frac{(\frac{t-1+h}{2})!}{h!(\frac{t-1-h}{2})!}(\cos\theta)^{h} \sin \theta  \begin{pmatrix}
     h \\
     g \\
 \end{pmatrix}\chi^{\frac{t-1+h-2g}{2}}e^{i\phi_2}\alpha_{x'} \int_{-\pi}^{\pi}\frac{dk}{2\pi}e^{ik(h-2g+1+x'-x)}\\
        & - \sum_{h=0}^{t-1} \sum_{g=0}^{h} \sum_{x'}(-1)^{h-g}\frac{(\frac{t-1+h}{2})!}{h!(\frac{t-1-h}{2})!}(\cos\theta)^{h+1} \begin{pmatrix}
     h \\
     g \\
 \end{pmatrix}\chi^{\frac{t-1+h-2g}{2}} \beta_{x'} \int_{-\pi}^{\pi}\frac{dk}{2\pi}e^{ik(h-2g-1+x'-x)}
\end{split}
\end{equation}
where $\chi=e^{i(\phi_1 +\phi_2)}$. Using $\int_{-\pi}^{\pi}\frac{dk}{2\pi}e^{ik(x-a)}=\delta(x-a)$, we have the conditions of $\delta(h-2g+x'-x) \implies g=\frac{h+x'-x}{2}$, $\delta(h-2g+1+x'-x) \implies g=\frac{h+1+x'-x}{2}$ and $\delta(h-2g-1+x'-x) \implies g=\frac{h-1+x'-x}{2}$. Then, from the properties of the delta function, we can argue that only a single term in the summation over $g$ will remain in each of the sums given in \eqref{Eq 2}. Hence, the explicit forms of coin coefficients in $x$-basis can be written as in \eqref{coin coeff in x-basis for localized state}. 

\section{Powers of the ${\mathcal{L}}_{k,k'}$ operator  } \label{Power of L_kk}
\noindent Considering the Hadamard coin given in \eqref{Hadmard coin}, the super-operator ${\mathcal{L}}_{k,k'}$ corresponding to the mixed localized initial states can be written as
\begin{equation}
   {\mathcal{L}}_{k,k'}= \left(
\begin{array}{cccc}
 \cos (k-k') & -i \sin (k-k') & 0 & 0 \\
 0 & 0 & \sin (k+k') & \cos (k+k') \\
 0 & 0 & -\cos (k+k') & \sin (k+k') \\
 -i \sin (k-k') & \cos (k-k') & 0 & 0 \\
\end{array}
\right)
\end{equation}
The characteristic equation of ${\mathcal{L}}_{k,k'}$ matrix is given by
\begin{equation}\label{charactersitic Eq L_kk}
    R(\lambda)=\lambda^4-c_0\lambda^3-c_1\lambda^2-c_2\lambda-c_3
\end{equation}
where $c_0 = \cos{(k - k')}-\cos{(k + k')} \\
c_1 = 2\cos{(k - k')}\cos{(k + k')} \\
c_2 = \cos{(k - k')}-\cos{(k + k')} \\
c_3=-1 \\$
\noindent Then, using the Fibonacci-Horner power decomposition, we can write the $t^{th}$ power of ${\mathcal{L}}_{k,k'}$ as
\begin{equation}\label{power_decomposition of L_kk}
    \mathcal{L}_{k,k'}^t=f_{t}\mathcal{L}_0+f_{t-1}\mathcal{L}_1+ f_{t-2}\mathcal{L}_2+f_{t-3}\mathcal{L}_3
\end{equation}
such that 
\begin{equation*}
    \mathcal{L}_0=\mathbb{I}_{4 \times 4}
\end{equation*}

\begin{equation*}
\begin{split}
    \mathcal{L}_1&=\mathcal{L}_{k,k'}-c_0\mathbb{I}_{4 \times 4} \\
    &=\left(
\begin{array}{cccc}
 \cos (k+k) & -i \sin (k-k') & 0 & 0 \\
 0 & -2 \sin k \sin k' & \sin (k+k') & \cos (k+k') \\
 0 & 0 & -\cos (k-k') & \sin (k+k') \\
 -i \sin (k-k') & \cos (k-k') & 0 & -2 \sin k \sin k' \\
\end{array}
\right)
\end{split}
\end{equation*}

\begin{equation*}
\begin{split}
    \mathcal{L}_2&=\mathcal{L}_{k,k'}^2-c_0\mathcal{L}_{k,k'}-c_1\mathbb{I}_{4 \times 4} \\
    &=\left(
\begin{array}{cccc}
 -\cos (k-k') \cos (k+k') & -i \cos (k+k') \sin (k-k') & -i \sin (k-k') \sin (k+k') & -i \cos (k+k') \sin (k-k') \\
 -i \cos (k+k') \sin (k-k') & -\cos (k-k') \cos (k+k') & -\cos (k-k') \sin (k+k') & -\frac{1}{2} (\cos 2 k+\cos 2 k'- 2) \\
 -i \sin (k-k') \sin (k+k') & \cos (k-k') \sin (k+k') & -\cos (k-k') \cos (k+k') & -\cos (k-k') \sin (k+k') \\
 -i \cos (k+k') \sin (k-k') & \frac{1}{2} (\cos 2 k+\cos 2 k'-2) & \cos (k-k') \sin (k+k') & -\cos (k-k') \cos (k+k') \\
\end{array}
\right)
\end{split}
\end{equation*}

\begin{equation*}
\begin{split}
    \mathcal{L}_3&=\mathcal{L}_{k,k'}^3-c_0\mathcal{L}_{k,k'}^2-c_1\mathcal{L}_{k,k'}-c_2\mathbb{I}_{4 \times 4}\\
    &=\left(
\begin{array}{cccc}
 -\cos (k-k') & 0 & 0 & -i \sin (k-k') \\
 -i \sin (k-k') & 0 & 0 & -\cos (k-k') \\
 0 & -\sin (k+k') & \cos (k+k') & 0 \\
 0 & -\cos (k+k') & -\sin (k+k') & 0 \\
\end{array}
\right)
\end{split}
\end{equation*}
where $c_i$s are the coefficients of the characteristic equation in \eqref{charactersitic Eq L_kk} and the function $f_{t}$ is defined by 
\begin{equation}
    f_t=\sum_{h_0+2h_1+3h_2+4h_3=t} \frac{(h_0+h_1+h_2+h_3)!}{h_0!h_1!h_2!h_3!}c_0^{h_0}c_1^{h_1} c_2^{h_2} c_3^{h_3} 
\end{equation}
\noindent for every $t \geq 1$. The set of matrices $\{\mathcal{L}_0, \mathcal{L}_1,\mathcal{L}_2,\mathcal{L}_3\}$ is called the Fibonacci-Horner basis of the power decomposition of $\mathcal{L}_{k,k'}$ \cite{laarichi2022explicit}. By applying the binomial expansion to the powers of the $c_i$ values given in \eqref{charactersitic Eq L_kk}, we can write the following expressions
\begin{equation}\label{powers of the coeff of the charactersitic poly of L_E}
\begin{split}
c_0^{h_0}&=\sum_{s_0=0}^{h_0}(-1)^{h_0-s_0}     
        \begin{pmatrix} h_0 \\ s_0\\
        \end{pmatrix}\bigg[\cos(k-k')\bigg]^{s_0}\bigg[\cos(k+k')\bigg]^{h_0-s_0}\\
c_1^{h_1}&=2^{h_1}\bigg[\cos(k-k')\bigg]^{h_1} 
        \bigg[\cos(k+k')\bigg]^{h_1}\\
c_2^{h_2}&=\sum_{s_2=0}^{h_2}(-1)^{h_2-s_2}     
        \begin{pmatrix} h_2 \\ s_2\\
        \end{pmatrix}\bigg[\cos(k-k')\bigg]^{s_2}\bigg[\cos(k+k')\bigg]^{h_2-s_2}\\
c_3^{h_3}&=(-1)^{h_3}
\end{split}
\end{equation}
where $\begin{pmatrix} n \\ m\\
\end{pmatrix}=\frac{n!}{m!(n-m)!}$ is the binomial coefficient. Write $\overline{h} = h_0+2h_1+3h_2+4h_3$. Then, the function $f_t$ can be rewritten in the following way
\begin{equation}\label{function f_t of L_kk}
\begin{split}
         f_t=\sum_{\overline{h}=t} \sum_{s_0=0}^{h_0}\sum_{s_2=0}^{h_2} 2^{h_1} \bigg[(-1)^{h_0+h_2+h_3-s_0-s_2}\bigg] &\bigg[\frac{(h_0+h_1+h_2+h_3)!}{h_0!h_1!h_2!h_3!}\bigg]\begin{pmatrix} h_0 \\ s_0\\ \end{pmatrix} \begin{pmatrix} h_2 \\ s_2\\ \end{pmatrix} \\ 
         & \times \bigg[\cos(k-k')\bigg]^{A_1}\bigg[\cos(k+k')\bigg]^{A_2}
\end{split}
\end{equation}
where the exponents of cosine function are given by $A_1=h_1+s_0+s_2$ and $A_2=h_0+h_1+h_2-s_0-s_2$. Then, by expressing the cosine in terms of complex exponential function and applying the binomial theorem again, we can write the following expression
\begin{equation}\label{function f_t of L_kk 2}
\begin{split}
    \bigg[\cos(k-k')\bigg]^{A_1}\bigg[\cos(k+k')\bigg]^{A_2} & =\sum_{v_1=0}^{A_1} \sum_{v_2=0}^{A_2} \bigg(\frac{1}{2}\bigg)^{A_1+A_2}\begin{pmatrix} A_1 \\ v_1\\ \end{pmatrix} \begin{pmatrix} A_2 \\ v_2\\ \end{pmatrix} \times e^{ik[2v_1+2v_2-A_1-A_2]}e^{ik'[2v_2-2v_1+A_1-A_2]}
\end{split}
\end{equation}
Now we can separate the $k$ and $k'$ variables and write $f_t$ in terms of exponential functions as follows
\begin{equation}\label{function f_t of L_kk 2}
     f_t=\sum_{\overline{h}=t} \sum_{s_0=0}^{h_0}\sum_{s_2=0}^{h_2} \sum_{v_1=0}^{A_1} \sum_{v_2=0}^{A_2} A_0 \begin{pmatrix} A_1 \\ v_1\\ \end{pmatrix} \begin{pmatrix} A_2 \\ v_2\\ \end{pmatrix} \times e^{ik[2v_1+2v_2-A_1-A_2]}e^{ik'[2v_2-2v_1+A_1-A_2]}
\end{equation}
where $A_0=\frac{2^{h_1}}{2^{(A_1+A_2)} }\bigg[(-1)^{h_0+h_2+h_3-s_0-s_2}\bigg]\bigg[\frac{(h_0+h_1+h_2+h_3)!}{h_0!h_1!h_2!h_3!}\bigg]\begin{pmatrix} h_0 \\ s_0\\ \end{pmatrix} \begin{pmatrix} h_2 \\ s_2\\ \end{pmatrix}$ and the exponents are given by $A_1=h_1+s_0+s_2$ and $A_2=h_0+h_1+h_2-s_0-s_2$. \\

\section{Derivation of the probability distribution related to the mixed localized initial state } \label{Prob_dis_of_mixed_localized_state}
\noindent Let us find an analytical expression for the probability distribution of the Hadamard walk that initiates for the mixed localized state given in \eqref{mixed initial state}. By substituting the expression \eqref{power_decomposition of L_kk} in \eqref{sup_Operator} and taking out the trace, we can determine an analytical expression for the trace of the operator  $\mathcal{L}_{k,k'}^{t}\mathbb{O}$ as
\begin{equation}\label{trace of the operaotr (L_k,k')^tO}
\begin{split}
    Tr(\mathcal{L}_{k,k'}^{t}\mathbb{O})=&2 r_0 f_t+f_{t-1} \bigg[2 r_0 \cos (k+k')-2 i r_2 \sin (k-k')\bigg]\\
    &-f_{t-2} \bigg[2 i r_1 \sin (k-k') \sin (k+k')+2r_0 \cos (k-k')\cos (k+k') + 2 i (r_2+r_3) \sin (k-k') \cos (k+k')\bigg] \\
    &-f_{t-3} \bigg[2 r_0 \cos (k-k')+2 i r_3 \sin (k-k')\bigg]
\end{split}
\end{equation}
Let us substitute \eqref{trace of the operaotr (L_k,k')^tO} in \eqref{position_probability} and write an expression for the probability distribution of the Hadamard walk that initiates from the mixed localized state as follows
\begin{equation}\label{position_probability2}
\begin{split}
    &P_{l}(y,t)=\int{\frac{dk}{2\pi }} \int{\frac{dk'}{2\pi }}e^{iy(k-k')}2 r_0 f_t + \int {\frac{dk}{2\pi }} \int {\frac{dk'}{2\pi }}e^{iy(k-k')} f_{t-1} \bigg[2 r_0 \cos (k+k')-2 i r_2 \sin (k-k')\bigg]\\
    &- \int {\frac{dk}{2\pi }} \int {\frac{dk'}{2\pi }}e^{iy(k-k')} f_{t-2} \bigg[2 i r_1 \sin (k-k') \sin (k+k')+2r_0 \cos (k-k')\cos (k+k') + 2 i (r_2+r_3) \sin (k-k') \cos (k+k')\bigg] \\
    &-\int {\frac{dk}{2\pi }} \int {\frac{dk'}{2\pi }}e^{iy(k-k')}f_{t-3} \bigg[2 r_0 \cos (k-k')+2 i r_3 \sin (k-k')\bigg]
\end{split}
\end{equation} 
where all the integrals in $P_{l}(y,t)$ are taken from $-\pi$ to $\pi$. As the final step, we need to evaluate the integrals over $k$ and $k'$ in \eqref{position_probability2}. Let us integrate terms containing $f_t$, $f_{t-1}$, $f_{t-2}$ and $f_{t-3}$ separately. To evaluate these integrals we substitute the function $f_t$ in \eqref{function f_t of L_kk 2} and make use of the results given in \eqref{set of intergrals}. Note that since the delta function appears in the each integration, when evaluating $P_{l}(y,t)$, the summations over $v_1$ and $v_2$ in function $f_t$ vanish in each case. The following set of integrals are used to compute the expression in \eqref{position_probability2}.

\begin{equation}\label{set of intergrals}
    \begin{split}
        &\int_{-\pi}^{\pi}\frac{dk}{2\pi}\int_{-\pi}^{\pi}\frac{dk'}{2\pi}e^{ikA}e^{ik'B}\cos(k+k')=\frac{1}{2}\delta(A+1)\delta(B+1)+\frac{1}{2}\delta(A-1)\delta(B-1) \\
         &\int_{-\pi}^{\pi}\frac{dk}{2\pi}\int_{-\pi}^{\pi}\frac{dk'}{2\pi}e^{ikA}e^{ik'B}\cos(k-k')=\frac{1}{2}\delta(A+1)\delta(B-1)+\frac{1}{2}\delta(A-1)\delta(B+1) \\
         &\int_{-\pi}^{\pi}\frac{dk}{2\pi}\int_{-\pi}^{\pi}\frac{dk'}{2\pi}e^{ikA}e^{ik'B}\sin(k-k')=\frac{1}{2i}\delta(A+1)\delta(B-1)-\frac{1}{2i}\delta(A-1)\delta(B+1) \\
        & \int_{-\pi}^{\pi}\frac{dk}{2\pi}\int_{-\pi}^{\pi}\frac{dk'}{2\pi}e^{ikA}e^{ik'B}\cos(k+k')\cos(k-k')=\frac{1}{4}\delta(A+2)\delta(B)+\frac{1}{4}\delta(A)\delta(B+2)+\frac{1}{4}\delta(A)\delta(B-2)+\frac{1}{4}\delta(A-2)\delta(B) \\
        & \int_{-\pi}^{\pi}\frac{dk}{2\pi}\int_{-\pi}^{\pi}\frac{dk'}{2\pi}e^{ikA}e^{ik'B}\sin(k+k')\sin(k-k')=\frac{1}{4}\delta(A)\delta(B-2)+\frac{1}{4}\delta(A)\delta(B+2)-\frac{1}{4}\delta(A+2)\delta(B)-\frac{1}{4}\delta(A-2)\delta(B) \\
        & \int_{-\pi}^{\pi}\frac{dk}{2\pi}\int_{-\pi}^{\pi}\frac{dk'}{2\pi}e^{ikA}e^{ik'B}\cos(k+k')\sin(k-k')=\frac{1}{4i}\delta(A+2)\delta(B)-\frac{1}{4i}\delta(A)\delta(B+2)+\frac{1}{4i}\delta(A)\delta(B-2)-\frac{1}{4i}\delta(A-2)\delta(B) \\
    \end{split}
\end{equation}
\subsection{Terms containing $f_t$}
\begin{equation}\label{Int Eq 1}
    \begin{split}
    &\int{\frac{dk}{2\pi }} \int{\frac{dk'}{2\pi }}e^{iy(k-k')}2 r_0 f_t \\
    & = 2r_0\sum A_0 \begin{pmatrix} A_1 \\ v_1\\ \end{pmatrix} \begin{pmatrix} A_2 \\ v_2\\ \end{pmatrix} \\
    & \times \int{\frac{dk}{2\pi }} \int{\frac{dk'}{2\pi }} e^{ik[2v_1+2v_2-A_1-A_2+y]}e^{ik'[2v_2-2v_1+A_1-A_2-y]} \\
    & = 2r_0 \sum_{\overline{h}=t} \sum_{s_0=0}^{h_0}\sum_{s_2=0}^{h_2} A_0 \begin{pmatrix} A_1 \\ \frac{A_1-y}{2}\\ \end{pmatrix} \begin{pmatrix} A_2 \\ \frac{A_2}{2}\\ \end{pmatrix}
\end{split} 
\end{equation}

\subsection{Terms containing $f_{t-1}$}

\begin{equation}\label{Int Eq 2}
    \begin{split}
    & \int{\frac{dk}{2\pi }} \int{\frac{dk'}{2\pi }}e^{iy(k-k')}2 r_0 f_{t-1}\cos(k+k') \\
    & = \ 2r_0\sum A_0 \begin{pmatrix} A_1 \\ v_1\\ \end{pmatrix} \begin{pmatrix} A_2 \\ v_2\\ \end{pmatrix} \\
    \times &\int{\frac{dk}{2\pi }} \int{\frac{dk'}{2\pi }} e^{ik[2v_1+2v_2-A_1-A_2+y]}e^{ik'[2v_2-2v_1+A_1-A_2-y]}\cos(k+k')  \\
    & = \ 2r_0 \sum_{\overline{h}=t} \sum_{s_0=0}^{h_0}\sum_{s_2=0}^{h_2} A_0 \bigg[\frac{1}{2}\begin{pmatrix} A_1 \\ \frac{A_1-y}{2}\\ \end{pmatrix} \begin{pmatrix} A_2 \\ \frac{A_2-1}{2}\\ \end{pmatrix} + \frac{1}{2} \begin{pmatrix} A_1 \\ \frac{A_1-y}{2}\\ \end{pmatrix} \begin{pmatrix} A_2 \\ \frac{A_2+1}{2}\\ \end{pmatrix} \bigg] \\
    & = \ 2r_0 \sum_{\overline{h}=t} \sum_{s_0=0}^{h_0}\sum_{s_2=0}^{h_2} A_0 \begin{pmatrix} A_1 \\ \frac{A_1-y}{2}\\ \end{pmatrix} \begin{pmatrix} A_2 \\ \frac{A_2-1}{2}\\ \end{pmatrix} 
\end{split} 
\end{equation}

\begin{equation}\label{Int Eq 3}
    \begin{split}
    & \int{\frac{dk}{2\pi }} \int{\frac{dk'}{2\pi }}e^{iy(k-k')}2 i r_2 f_{t-1}\sin(k-k') \\
    & = \ 2 i r_2\sum A_0 \begin{pmatrix} A_1 \\ v_1\\ \end{pmatrix} \begin{pmatrix} A_2 \\ v_2\\ \end{pmatrix} \\
    & \times \int{\frac{dk}{2\pi }} \int{\frac{dk'}{2\pi }} e^{ik[2v_1+2v_2-A_1-A_2+y]}e^{ik'[2v_2-2v_1+A_1-A_2-y]}\sin(k-k')  \\
    & = \ 2 i r_2 \sum_{\overline{h}=t} \sum_{s_0=0}^{h_0}\sum_{s_2=0}^{h_2} A_0 \bigg[\frac{1}{2i}\begin{pmatrix} A_1 \\ \frac{A_1-y-1}{2}\\ \end{pmatrix} \begin{pmatrix} A_2 \\ \frac{A_2}{2}\\ \end{pmatrix} - \frac{1}{2i}\begin{pmatrix} A_1 \\ \frac{A_1-y+1}{2}\\ \end{pmatrix} \begin{pmatrix} A_2 \\ \frac{A_2}{2}\\ \end{pmatrix} \bigg] \\
    & = \ r_2 \sum_{\overline{h}=t} \sum_{s_0=0}^{h_0}\sum_{s_2=0}^{h_2}  \frac{-yA_0}{A_1+1}\begin{pmatrix} A_1+1 \\ \frac{A_1-y+1}{2}\\ \end{pmatrix} \begin{pmatrix} A_2 \\ \frac{A_2}{2}\\ \end{pmatrix} 
\end{split} 
\end{equation}
Note that in the third step of the integral in \eqref{Int Eq 2}, we use the fact that $\begin{pmatrix} n \\ m\\ \end{pmatrix} = \begin{pmatrix} n \\ n-m\\ \end{pmatrix}$ and take $n=A_2$ and $m=\frac{A_2-1}{2}$. Moreover, in the third step of the integral in \eqref{Int Eq 3}, we use $\begin{pmatrix} n \\ m-1\\ \end{pmatrix} - \begin{pmatrix} n \\ m\\ \end{pmatrix} =\frac{2m-n-1}{n+1}\begin{pmatrix} n +1\\ m\\ \end{pmatrix} $ and take $n=A_1$ and $m=\frac{A_1-y+1}{2}$. 

\subsection{Terms containing $f_{t-2}$}

\begin{equation}\label{Int Eq 4}
    \begin{split}
    & \int{\frac{dk}{2\pi }} \int{\frac{dk'}{2\pi }}e^{iy(k-k')}2 i r_1 f_{t-2}\sin(k+k') \sin(k-k') \\
    & = 2 i r_1\sum A_0 \begin{pmatrix} A_1 \\ v_1\\ \end{pmatrix} \begin{pmatrix} A_2 \\ v_2\\ \end{pmatrix} \\
    & \times \int{\frac{dk}{2\pi }} \int{\frac{dk'}{2\pi }} e^{ik[2v_1+2v_2-A_1-A_2+y]}e^{ik'[2v_2-2v_1+A_1-A_2-y]}\sin(k+k') \sin(k-k')   \\
    & = \ 2 i r_1 \sum_{\overline{h}=t} \sum_{s_0=0}^{h_0}\sum_{s_2=0}^{h_2} A_0 \bigg[\frac{1}{4}\begin{pmatrix} A_1 \\ \frac{A_1-y-1}{2}\\ \end{pmatrix} \begin{pmatrix} A_2 \\ \frac{A_2+1}{2}\\ \end{pmatrix} + \frac{1}{4}\begin{pmatrix} A_1 \\ \frac{A_1-y+1}{2}\\ \end{pmatrix} \begin{pmatrix} A_2 \\ \frac{A_2-1}{2}\\ \end{pmatrix} \\
    & - \frac{1}{4}\begin{pmatrix} A_1 \\ \frac{A_1-y-1}{2}\\ \end{pmatrix} \begin{pmatrix} A_2 \\ \frac{A_2-1}{2}\\ \end{pmatrix}-\frac{1}{4}\begin{pmatrix} A_1 \\ \frac{A_1-y+1}{2}\\ \end{pmatrix} \begin{pmatrix} A_2 \\ \frac{A_2+1}{2}\\ \end{pmatrix}\bigg] \\
    & =  0
\end{split} 
\end{equation}

\begin{equation}\label{Int Eq 5}
    \begin{split}
    & \int{\frac{dk}{2\pi }} \int{\frac{dk'}{2\pi }}e^{iy(k-k')}2 r_0 f_{t-2}\cos(k+k') \cos(k-k') \\ 
    & = 2 r_0\sum A_0 \begin{pmatrix} A_1 \\ v_1\\ \end{pmatrix} \begin{pmatrix} A_2 \\ v_2\\ \end{pmatrix} \\
    & \times \int{\frac{dk}{2\pi }} \int{\frac{dk'}{2\pi }} e^{ik[2v_1+2v_2-A_1-A_2+y]}e^{ik'[2v_2-2v_1+A_1-A_2-y]}\cos(k+k') \cos(k-k')   \\
    & = \ 2 r_0 \sum_{\overline{h}=t} \sum_{s_0=0}^{h_0}\sum_{s_2=0}^{h_2} A_0 \bigg[\frac{1}{4}\begin{pmatrix} A_1 \\ \frac{A_1-y-1}{2}\\ \end{pmatrix} \begin{pmatrix} A_2 \\ \frac{A_2-1}{2}\\ \end{pmatrix} + \frac{1}{4}\begin{pmatrix} A_1 \\ \frac{A_1-y+1}{2}\\ \end{pmatrix} \begin{pmatrix} A_2 \\ \frac{A_2-1}{2}\\ \end{pmatrix} \\
    & + \frac{1}{4}\begin{pmatrix} A_1 \\ \frac{A_1-y-1}{2}\\ \end{pmatrix} \begin{pmatrix} A_2 \\ \frac{A_2+1}{2}\\ \end{pmatrix}+\frac{1}{4}\begin{pmatrix} A_1 \\ \frac{A_1-y+1}{2}\\ \end{pmatrix} \begin{pmatrix} A_2 \\ \frac{A_2+1}{2}\\ \end{pmatrix}\bigg] \\
    & = \ 2 r_0 \sum_{\overline{h}=t} \sum_{s_0=0}^{h_0}\sum_{s_2=0}^{h_2} A_0 \bigg[\frac{1}{2}\begin{pmatrix} A_1 \\ \frac{A_1-y-1}{2}\\ \end{pmatrix} \begin{pmatrix} A_2 \\ \frac{A_2-1}{2}\\ \end{pmatrix} + \frac{1}{2}\begin{pmatrix} A_1 \\ \frac{A_1-y+1}{2}\\ \end{pmatrix} \begin{pmatrix} A_2 \\ \frac{A_2-1}{2}\\ \end{pmatrix} \bigg] \\
    & = \ r_0 \sum_{\overline{h}=t} \sum_{s_0=0}^{h_0}\sum_{s_2=0}^{h_2} A_0 \begin{pmatrix} A_1+1 \\ \frac{A_1-y+1}{2}\\ \end{pmatrix} \begin{pmatrix} A_2 \\ \frac{A_2-1}{2}\\ \end{pmatrix} 
\end{split} 
\end{equation}

\begin{equation}\label{Int Eq 6}
    \begin{split}
    & \int{\frac{dk}{2\pi }} \int{\frac{dk'}{2\pi }}e^{iy(k-k')}2 i (r_2+r_3) f_{t-2}\cos(k+k')  \sin(k-k') \\
    & = 2 i (r_2+r_3) \sum A_0 \begin{pmatrix} A_1 \\ v_1\\ \end{pmatrix} \begin{pmatrix} A_2 \\ v_2\\ \end{pmatrix} \\
    & \times \int{\frac{dk}{2\pi }} \int{\frac{dk'}{2\pi }} e^{ik[2v_1+2v_2-A_1-A_2+y]}e^{ik'[2v_2-2v_1+A_1-A_2-y]}\cos(k+k') \sin(k-k')   \\
    & = \ 2 i (r_2+r_3) \sum_{\overline{h}=t} \sum_{s_0=0}^{h_0}\sum_{s_2=0}^{h_2} A_0 \bigg[\frac{1}{4i}\begin{pmatrix} A_1 \\ \frac{A_1-y-1}{2}\\ \end{pmatrix} \begin{pmatrix} A_2 \\ \frac{A_2-1}{2}\\ \end{pmatrix} - \frac{1}{4i}\begin{pmatrix} A_1 \\ \frac{A_1-y+1}{2}\\ \end{pmatrix} \begin{pmatrix} A_2 \\ \frac{A_2-1}{2}\\ \end{pmatrix} \\
    & + \frac{1}{4i}\begin{pmatrix} A_1 \\ \frac{A_1-y-1}{2}\\ \end{pmatrix} \begin{pmatrix} A_2 \\ \frac{A_2+1}{2}\\ \end{pmatrix}-\frac{1}{4i}\begin{pmatrix} A_1 \\ \frac{A_1-y+1}{2}\\ \end{pmatrix} \begin{pmatrix} A_2 \\ \frac{A_2+1}{2}\\ \end{pmatrix}\bigg] \\
    & = 0
\end{split} 
\end{equation}
Note that in the third step of the integrals in \eqref{Int Eq 4} and \eqref{Int Eq 6} , we use the fact that $\begin{pmatrix} n \\ m\\ \end{pmatrix} = \begin{pmatrix} n \\ n-m\\ \end{pmatrix}$ and take $n=A_2$ and $m=\frac{A_2-1}{2}$. Moreover, in the third step of the integral in \eqref{Int Eq 5}, we use the identity $\begin{pmatrix} n +1\\ m\\ \end{pmatrix}=\begin{pmatrix} n \\ m-1\\ \end{pmatrix} + \begin{pmatrix} n \\ m\\ \end{pmatrix}$ and take $n=A_1$ and $m=\frac{A_1-y+1}{2}$.

\subsection{Terms containing $f_{t-3}$}

\begin{equation}\label{Int Eq 7}
    \begin{split}
       & \int{\frac{dk}{2\pi }} \int{\frac{dk'}{2\pi }}e^{iy(k-k')}2 r_0 f_{t-3}\cos(k-k') \\
    &= \ 2r_0\sum A_0 \begin{pmatrix} A_1 \\ v_1\\ \end{pmatrix} \begin{pmatrix} A_2 \\ v_2\\ \end{pmatrix} \\
    & \times \int{\frac{dk}{2\pi }} \int{\frac{dk'}{2\pi }} e^{ik[2v_1+2v_2-A_1-A_2+y]}e^{ik'[2v_2-2v_1+A_1-A_2-y]}\cos(k-k')  \\
    & = \ 2r_0 \sum_{\overline{h}=t} \sum_{s_0=0}^{h_0}\sum_{s_2=0}^{h_2} A_0 \bigg[\frac{1}{2}\begin{pmatrix} A_1 \\ \frac{A_1-y-1}{2}\\ \end{pmatrix} \begin{pmatrix} A_2 \\ \frac{A_2}{2}\\ \end{pmatrix} + \frac{1}{2}\begin{pmatrix} A_1 \\ \frac{A_1-y+1}{2}\\ \end{pmatrix} \begin{pmatrix} A_2 \\ \frac{A_2}{2}\\ \end{pmatrix} \bigg] \\
    & = \ r_0 \sum_{\overline{h}=t} \sum_{s_0=0}^{h_0}\sum_{s_2=0}^{h_2} A_0 \begin{pmatrix} A_1+1 \\ \frac{A_1-y+1}{2}\\ \end{pmatrix} \begin{pmatrix} A_2 \\ \frac{A_2}{2}\\ \end{pmatrix} 
\end{split} 
\end{equation}

\begin{equation}\label{Int Eq 8}
    \begin{split}
    & \int{\frac{dk}{2\pi }} \int{\frac{dk'}{2\pi }}e^{iy(k-k')}2 i r_3 f_{t-3}\sin(k-k') \\
    & = \ 2 i r_3\sum A_0 \begin{pmatrix} A_1 \\ v_1\\ \end{pmatrix} \begin{pmatrix} A_2 \\ v_2\\ \end{pmatrix} \\
    & \times \int{\frac{dk}{2\pi }} \int{\frac{dk'}{2\pi }} e^{ik[2v_1+2v_2-A_1-A_2+y]}e^{ik'[2v_2-2v_1+A_1-A_2-y]}\sin(k-k')  \\
    & = \ 2 i r_3 \sum_{\overline{h}=t} \sum_{s_0=0}^{h_0}\sum_{s_2=0}^{h_2} A_0 \bigg[\frac{1}{2i}\begin{pmatrix} A_1 \\ \frac{A_1-y-1}{2}\\ \end{pmatrix} \begin{pmatrix} A_2 \\ \frac{A_2}{2}\\ \end{pmatrix} - \frac{1}{2i}\begin{pmatrix} A_1 \\ \frac{A_1-y+1}{2}\\ \end{pmatrix} \begin{pmatrix} A_2 \\ \frac{A_2}{2}\\ \end{pmatrix} \bigg] \\
    & = \ r_3 \sum_{\overline{h}=t} \sum_{s_0=0}^{h_0}\sum_{s_2=0}^{h_2}  \frac{-yA_0}{A_1+1}\begin{pmatrix} A_1+1 \\ \frac{A_1-y+1}{2}\\ \end{pmatrix} \begin{pmatrix} A_2 \\ \frac{A_2}{2}\\ \end{pmatrix} 
\end{split} 
\end{equation}
Note that in the third step of the integral in \eqref{Int Eq 7}, we use the identity $\begin{pmatrix} n +1\\ m\\ \end{pmatrix}=\begin{pmatrix} n \\ m-1\\ \end{pmatrix} + \begin{pmatrix} n \\ m\\ \end{pmatrix}$ and take $n=A_1$ and $m=\frac{A_1-y+1}{2}$. Moreover, in the third step of the integral in \eqref{Int Eq 8}, we use the identity involving the binomial coefficients $\begin{pmatrix} n \\ m-1\\ \end{pmatrix} - \begin{pmatrix} n \\ m\\ \end{pmatrix} =\frac{2m-n-1}{n+1}\begin{pmatrix} n +1\\ m\\ \end{pmatrix} $ and take $n=A_1$ and $m=\frac{A_1-y+1}{2}$.




\end{appendices}


\bibliography{sn-bibliography}

\end{document}